\documentclass{article}
\usepackage{spconf,amsmath,graphicx}
\usepackage[ruled, vlined, linesnumbered]{algorithm2e}
\usepackage{bbm}
\usepackage{float}
\usepackage{multicol}
\usepackage{xcolor}
\usepackage[colorlinks = true,
            linkcolor = black,
            urlcolor  = blue,
            citecolor = black,
            anchorcolor = blue]{hyperref}

\title{Communication-Cost Aware Microphone Selection For Neural Speech Enhancement with Ad-hoc Microphone Arrays}
\name{Jonah Casebeer$^{\sharp}$ \qquad Jamshed Kaikaus$^{\sharp}$ \qquad Paris Smaragdis$^{\sharp,\flat}$}
\address{$^{\sharp}$ University of Illinois at Urbana-Champaign \\ $^{\flat}$ Adobe Research}
\begin{document}
%
\maketitle
\begin{abstract}
In this paper, we present a method for jointly-learning a microphone selection mechanism and a speech enhancement network for multi-channel speech enhancement with an ad-hoc microphone array. The attention-based microphone selection mechanism is trained to reduce communication-costs through a penalty term which represents a task-performance/ communication-cost trade-off. While working within the trade-off, our method can intelligently stream from more microphones in lower SNR scenes and fewer microphones in higher SNR scenes. We evaluate the model in complex echoic acoustic scenes with moving sources and show that it matches the performance of models that stream from a fixed number of microphones while reducing communication costs.
\end{abstract}
\begin{keywords}
sensor selection, ad-hoc microphone array, beamforming, speech enhancement, deep learning
\end{keywords}

\section{Introduction}
\label{sec:intro}
Complex auditory scenes may contain dozens of mobile devices with microphones such as smart-phones, smart-watches, laptops, etc. It is often desirable to leverage all microphones within such an ad-hoc array for tasks like multi-channel speech enhancement. This is of great interest thanks to applications such as hands-free teleconferencing. However, ad-hoc arrays present unique complications due to the communication expense of continuously streaming from many devices and their unruly geometry. 

When communication cost is not an issue, both beamforming and deep learning techniques have proven successful for multi-channel speech enhancement. These range from techniques that combine single-channel enhancers with beamformers \cite{erdogan2016improved, heymann2016neural} to fully learned models that leverage spatial features \cite{wang2018combining, wang2018multi}.

To reduce communication or computational costs, several general sensor selection algorithms have been derived \cite{joshi2008sensor, chepuri2014sparsity}. In the case of distributed acoustic sensors, additional techniques have been proposed to synchronize sensors \cite{cherkassky2017blind}, overcome the distributed nature of ad-hoc arrays \cite{heusdens2012distributed, zeng2013distributed, zhang2018rate, cherkassky2017blind}, and perform subset selection \cite{zhang2017microphone}. These algorithms typically define a task-performance vs communication-cost trade-off to leverage the fact that some microphones will not be necessary for good performance. By using a microphone selection algorithm in conjunction with a hand-tuned stopping criterion these algorithms decide whether to process data from a microphone without having seen it. In some cases, this selection step can be informed by sensor geometry.

However, these approaches are not amenable to being used with deep learning models due to the non-differentiability of selection. Motivated by this, we design a neural network that can learn, in a data-driven, fashion whether to process data from an unseen microphone, given a performance/cost trade-off. We circumvent non-differentiability by formulating a specialized attention-based mechanism that allows the model to learn a microphone selection mechanism jointly with a speech-enhancement model. Our mechanism is based on the adaptive compute technique explored in \cite{graves2016adaptive, DBLP:journals/corr/FigurnovCZZHVS16, DBLP:journals/corr/abs-1807-03819} and the multi-channel models presented in \cite{casebeer2018multi, casebeer2019multi}.

The novelty of our method is two-fold. First, using an attention-based mechanism for microphone selection and second, learning this mechanism jointly with the enhancement network. To evaluate our model, we construct a challenging multi-channel speech enhancement task using an ad-hoc array of microphones inside an echoic scene with moving sources. Our experiments show that the performance/cost trade-off can be easily selected and that the model works within the trade-off to learn an adaptive microphone selection criterion capable of requesting more microphones in more complex scenes and fewer microphone in simpler scenes. For baselines, we compare to models that select a fixed number of microphones. We use short-time objective intelligibility \cite{taal2011algorithm} to measure speech enhancement performance and seconds of audio streamed to measure communication costs. The code is available \href{https://jmcasebeer.github.io/projects/camse/}{\color{blue}{here}}.

\begin{figure*}[ht!]
    \centering
    \includegraphics[scale=0.72]{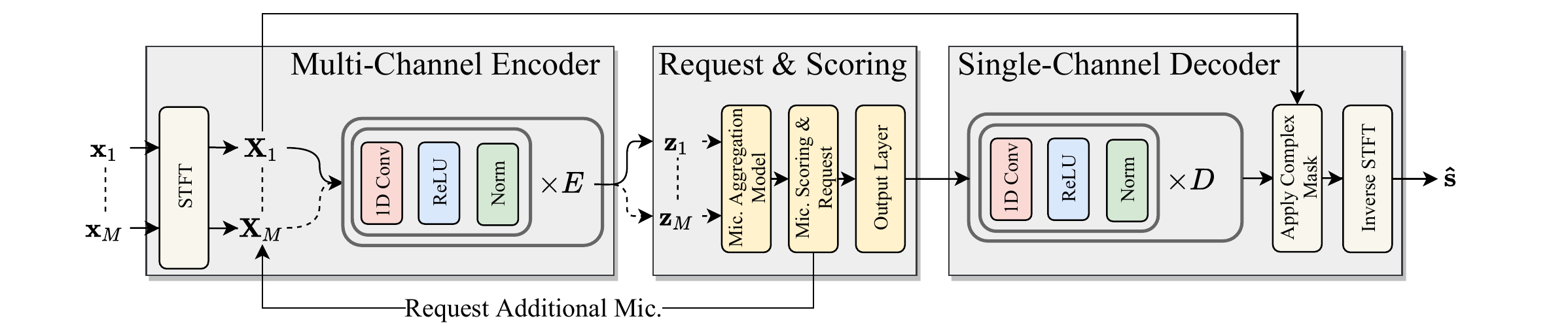}
    \caption{The model consists of an encoder, a request \& scoring mechanism, and a decoder. The encoder serves as a multi-channel feature extractor for the request \& scoring mechanism, which performs sequential microphone selection. The decoder then receives the aggregated microphone information and outputs a single-channel estimate of the clean speech.}
    \label{fig:model}
    \vspace{-2mm}
\end{figure*}

\vspace{-1mm}
\section{Methods}
\label{sec:methods}
\vspace{-1mm}

The task of multi-channel speech enhancement is to recover some clean speech signal $\mathbf{s}$ from a set of $M$ noisy reverberant mixtures $\mathbf{x}_i$. Each mixture is captured by a stationary microphone and is modeled by $\mathbf{x}_i = \mathbf{s}_i + \mathbf{n}_i, i \in \{1, \cdots, M\}$, where both $\mathbf{s}_i$ and $\mathbf{n}_i$ are the result of convolving a room impulse response with an anechoic signal. Thus, a multi-channel speech enhancement model takes as input $[\mathbf{x}_1, \cdots, \mathbf{x}_M]$ and returns an estimate of the clean speech, $\mathbf{\hat{s}}$. 

\subsection{Network Architecture}
\label{sec:model}
To develop a multi-channel speech enhancement model that is communication-cost aware, we propose a specialized microphone request mechanism. This mechanism allows the model to perform sequential microphone subset selection until a learned stopping criterion is satisfied. We construct this request mechanism by reinterpreting and repurposing the adaptive compute time mechanism presented in \cite{graves2016adaptive, DBLP:journals/corr/FigurnovCZZHVS16, DBLP:journals/corr/abs-1807-03819}. The proposed network architecture is depicted in Fig.~\ref{fig:model}. It consists of three components: an encoder which acts as a multi-channel feature extractor; a request \& scoring mechanism which aggregates microphone information; and a decoder, which produces a single-channel clean speech estimate.

\subsubsection{Encoder}
The multi-channel encoder operates on each microphone individually in order to perform feature extraction. It processes temporal chunks of single-channel mixtures as the request mechanism asks for them. It first computes the Short-time Fourier Transform~(STFT), and passes the magnitude through a series of $E$ 1D convolutional layers. For the $i^{th}$ microphone, the encoder outputs $\mathbf{z}[t]^{(i)}$ given the time-domain input chunk $\mathbf{x}_i[t \cdot h: t\cdot h + f]$. Here, $h$ represents the effective hop of all encoder convolutions and  $f$ represents the effective receptive-field. These constants are used in Alg.~\ref{alg:mic_request}.

\subsubsection{Request \& Scoring Mechanism}
The request \& scoring mechanism performs microphone selection by iteratively requesting to stream from microphones based on the aggregated sensor information. These modules are trained with a performance/cost trade-off that influences how many requests are made. Specifically, if the model has already processed $k$ channels, an attention layer aggregates across channels to produce an internal representation that is assigned a score. If the score is within the model's budget, then it will request data from an additional microphone. The score is order invariant with respect to the $k$ aggregated channels. Once the request budget is exhausted, the aggregated representation and scores are passed through a final layer to the decoder.

Aggregation is accomplished via a multi-head attention mechanism with internal dimension $d$, query $\mathbf{Q}[t]^{1 \times d}$, keys $\mathbf{K}[t]^{k \times d}$, and values $\mathbf{V}[t]^{k \times d}$. Unlike traditional self-attention, we use a single query computed from the most recently requested microphone to collect information across all previously requested microphones. The operations are given by $\mathbf{Q}[t] = \mathbf{W}_q \cdot \mathbf{z}[t]^{(k)}$, $\mathbf{K}[t] = \mathbf{W}_k \cdot \mathbf{z}[t]^{(0:k)}$, and $\mathbf{V}[t] = \mathbf{W}_v \cdot \mathbf{z}_[t]^{(0:k)}$  which are used to compute: $\mathbf{h}[t]^{(k)} = \text{Softmax}(\frac{\mathbf{Q}[t] \mathbf{K}[t]^\top}{\sqrt{d}}){\mathbf{V}[t]}$. This output represents aggregated information when using the $k$ requested microphones at time $t$. The second step is a position-wise feed-forward network used to further process $\mathbf{h}[t]^{(k)}$ \cite{vaswani2017attention}. Both steps are followed by dropout, a residual connection, and layer normalization. In Alg.~\ref{alg:mic_request} this entire set of operations is represented by $\text{AttentionLayer}(\mathbf{z}[t]^{(0:k)})$.

The request \& scoring module takes $\mathbf{h}[t]^{(k)}$, and calculates a score $s_k = S(\mathbf{h}[t]^{(k)})$ to determine if the model should request to stream data from another microphone. The scoring network outputs a scalar in the range $[0,1]$. As microphones are requested and scored, the scores are accumulated in the variable $c$. If $c$ is greater than the request budget $1 - \epsilon$, this signifies that we have seen enough data to cease requesting microphones. Otherwise, the model requests another microphone, computes $h[t]^{(k+1)}$, and assigns a new score. The scores are constrained to sum to $1$. If a score pushes the sum over one it is trimmed and referred to as the remainder $r$. Assuming $N$ microphones requests are made, the final hidden state is computed as $\mathbf{h}[t] = \sum_{i=1}^{N - 1} s_i \cdot \mathbf{h}[t]^{(i)} + r \cdot h[t]^{(N)}$.

Once the request budget has been exhausted and its remainder $r$ calculated, we pass $\mathbf{h}[t]$ through a dense layer, to obtain an input for the decoder. The value $p = N + r$ is saved for use in the loss function. Thus, this reformulation of adaptive compute attempts to examine as little data as possible by weighting an internal representation of the multi-channel scene. The complete algorithm is shown in Alg.~\ref{alg:mic_request}.

\begin{algorithm}[t!]
\SetAlgoLined
\KwResult{Speech Estimate $\mathbf{\hat{s}}$, request cost $p$}
 \textbf{Inputs:} $[\mathbf{x}_1, \cdots, \mathbf{x}_M], T, M, f, h$ \\
 \textbf{Hyperparameter:} $\epsilon$ \\
 $p=0, \mathbf{\hat{s}} = \mathbf{0}^{(T \times 1)}$\\
 \For{$t \gets 0$ to $\frac{T}{h}$} {
 $c=0; r=1; p_t=0$\\
 \For{$k \gets 1$ to $M$}{
  $\mathbf{z}[t]^{(k)} = \text{Encoder}(\mathbf{x}_k[t \cdot h : t \cdot h + f])$\\
  $\mathbf{h}[t]^{(k)} = \text{AttentionLayer}(\mathbf{z}[t]^{(0:k)})$ \\
  $s_{k}= S(\mathbf{h}[t]^{(k)})$\\
  \If{$k == M$}{
  $s_{k}= 1$
  }
  $c = c + s_{k}; p_t = p_t + 1$\\
  \eIf{$c < 1 - \epsilon$}{
  $\mathbf{h}[t] = \mathbf{h}[t] + s_{k} \cdot \mathbf{h}[t]^{(k)}$\\
  $r = r - s_{k}$\\
  }{
  $\mathbf{h}[t] = \mathbf{h}[t] + r \cdot \mathbf{h}[t]^{(k)}$\\
  $p_t = p_t + r$\\
  $\text{Break}$\\
  }
  }
  $\mathbf{\hat{s}}[t \cdot h: t \cdot h + f] = \mathbf{\hat{s}}[t \cdot h: t \cdot h + f] + \text{Decoder}(\mathbf{h[t]})$\\
  $p = p + p_t$
 }
\textbf{return} $\mathbf{\hat{s}}, \frac{p}{T / h}$\\

 \caption{Network Forward Pass}
 \label{alg:mic_request}
\end{algorithm}

\subsubsection{Decoder}

The decoder operates on the aggregated microphone information produced by the request mechanism. It predicts real and imaginary masks, denoted as $\mathbf{M}_r$ and $\mathbf{M}_i$, respectively, via a series of $D$ 1D convolutional layers. The masks are applied to the STFT computed in the encoder, to produce the clean speech estimate $\widehat{\mathbf{S}} = \mathbf{M}_r \odot \operatorname{Re}(\mathbf{X}) + \mathbf{M}_i \odot \operatorname{Im}(\mathbf{X})$. The STFT domain estimate $\widehat{\mathbf{S}}$ is converted to the time domain representation $\mathbf{\hat{s}}$ via the inverse STFT.

\subsubsection{Loss Function}
The full network is trained in an end-to-end fashion using two losses. The first loss $\mathcal{L}_p(p) = \sum_{t=0}^T p[t]$ encourages the model to request less data. Intuitively, since $p[t]$ represents the leftover request budget, minimizing it encourages the model to exhaust its budget sooner and thus request data from fewer microphones. The second loss measures speech enhancement performance between the true clean speech spectrogram $\mathbf{S}$ and the estimated spectrogram $\widehat{\mathbf{S}}$. It is defined as $\mathcal{L}_s(\mathbf{S}, \widehat{\mathbf{S}}) = \| |\widehat{\mathbf{S}}|^{\alpha}  - |\mathbf{S}|^{\alpha} \|_1$, where $\alpha=.3$ is applied element-wise and intended to provide more weight to low energy frequencies. The full loss is then $\mathcal{L} = \mathcal{L}_s(\mathbf{S}, \widehat{\mathbf{S}}) + \lambda \mathcal{L}_p(p)$, where $\lambda$ is a request penalty that specifies the performance/cost trade-off.

\subsection{Dataset Generation}
To generate multi-microphone scenes, we modified the Microsoft Scalable Noisy Speech Dataset \cite{reddy2019scalable} to use the \textit{pyroomacoustics} \cite{scheibler2018pyroomacoustics} implementation of the image source method. To simulate a scene, we select a speech sample and a noise sample, rescaling them to a random SNR in $[-10, 10]$ dB. The rescaled samples are placed in a simulated 3-D room with width, depth, and height distributed uniformly in the interval $[10, 15)$ meters, $10^{th}$ order reflections, and an absorption coefficient of $.35$. Within this room, microphone locations are distributed uniformly at random. Both the speech and noise source move with their starting and ending location also selected uniformly at random. We generate distinct testing and training sets which do not have any speech or noise overlap. All scenes last two seconds. 

\subsubsection{Baseline Model}
For comparison, we construct two baselines. The first baseline uses the same architecture as the proposed model but has request penalty $\lambda$ set to $0$ and therefore does not incur a penalty for requesting additional microphones. The second baseline has an identical encoder and decoder but does not perform adaptive data requests and instead processes a fixed number of microphones. All models have the same number of parameters.

\begin{figure*}[t]
\minipage{0.32\textwidth}
  \includegraphics[width=\linewidth]{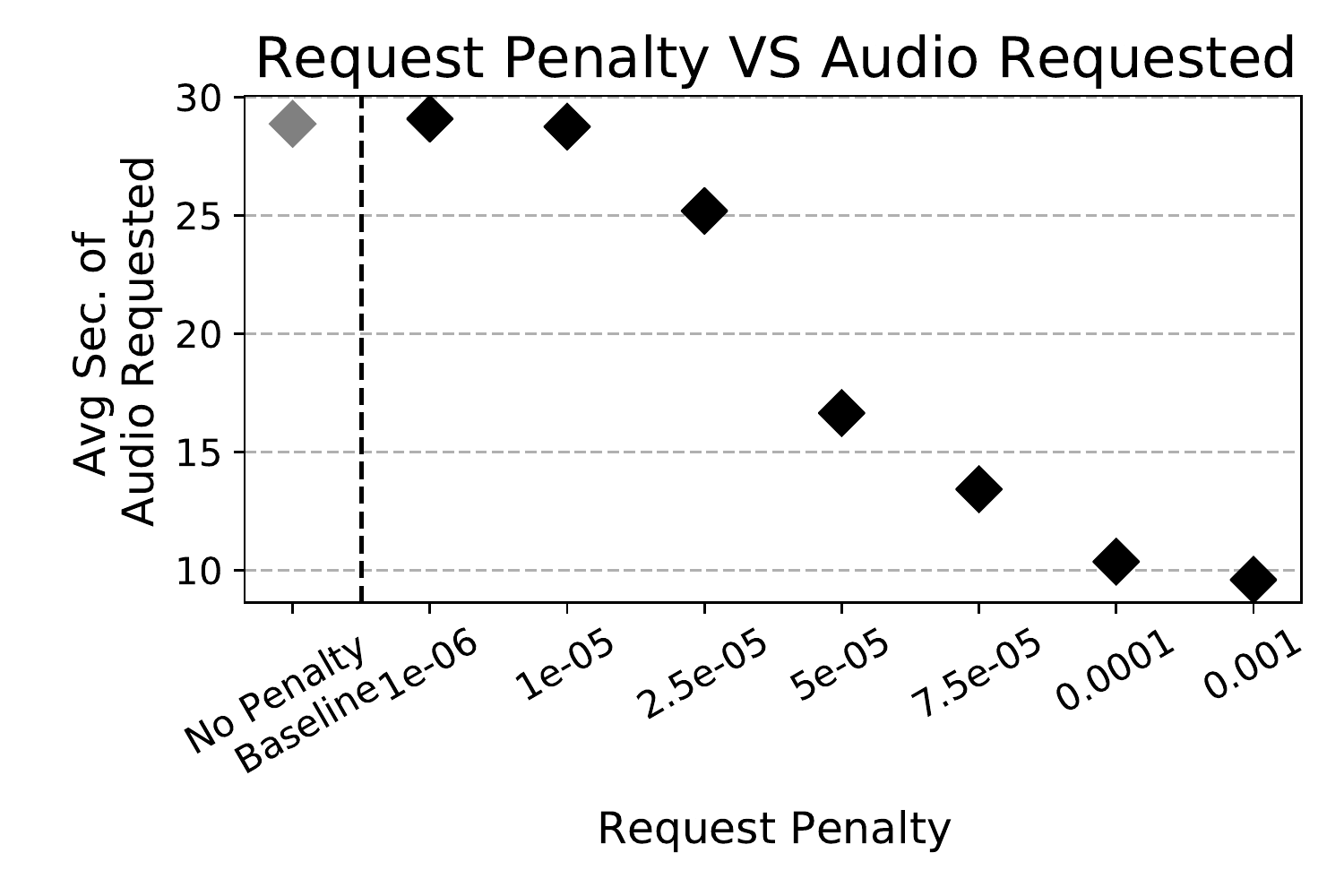}
  \caption{Effect of different request penalties on the average amount of data requested by our proposed models. As the request penalty increases, less data is requested.}
  \vspace{-1mm}
  \label{fig:tradeoff}
\endminipage\hfill
\minipage{0.32\textwidth}
  \includegraphics[width=\linewidth]{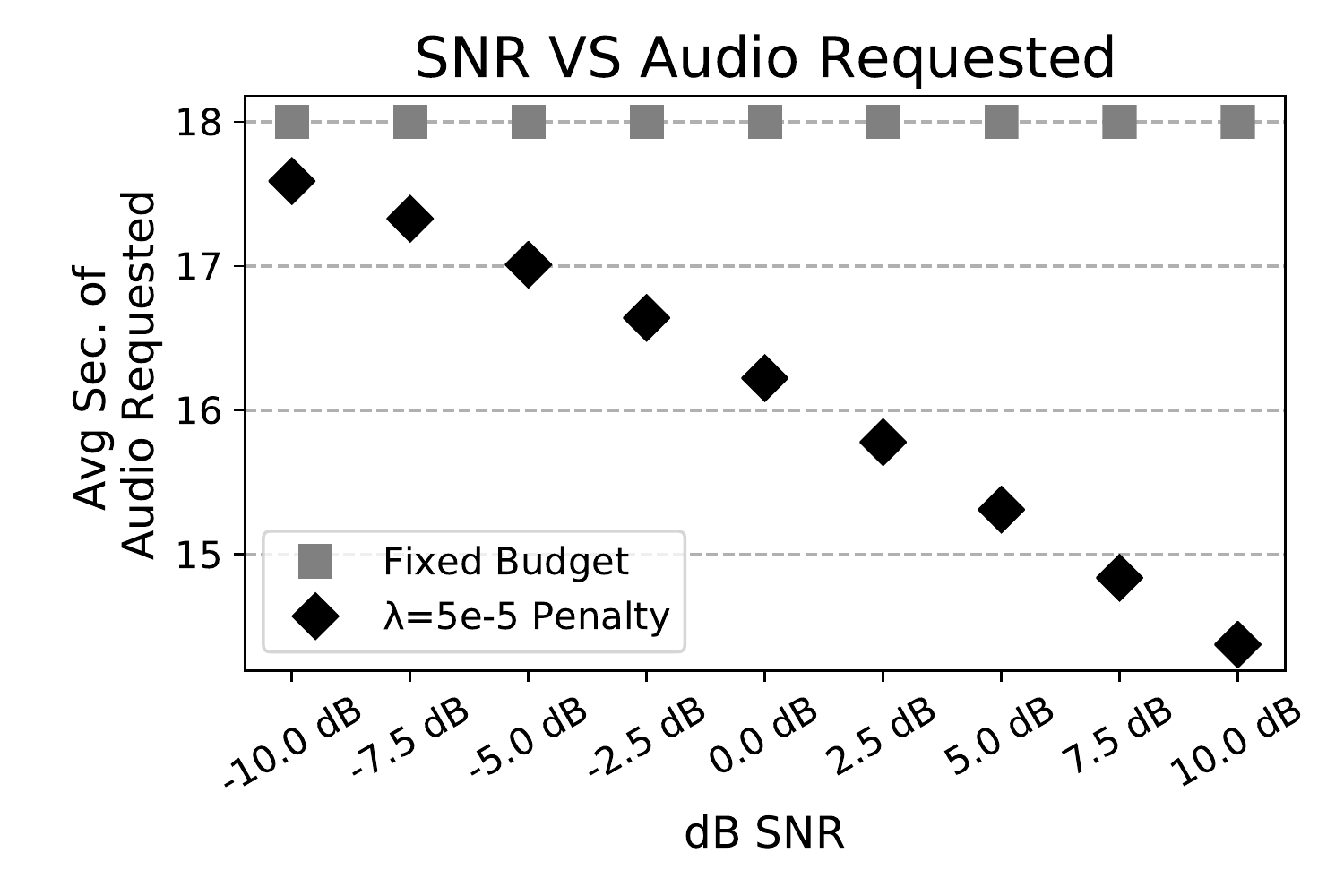}
  \caption{Amount of data requested by the proposed model~($\lambda=5\times10^{-5}$) compared to a model which requests a fixed amount of data. The proposed model adapts requests to the scene difficulty.}
  \vspace{-1mm}
  \label{fig:snr_sec}
\endminipage\hfill
\minipage{0.32\textwidth}%
  \includegraphics[width=\linewidth]{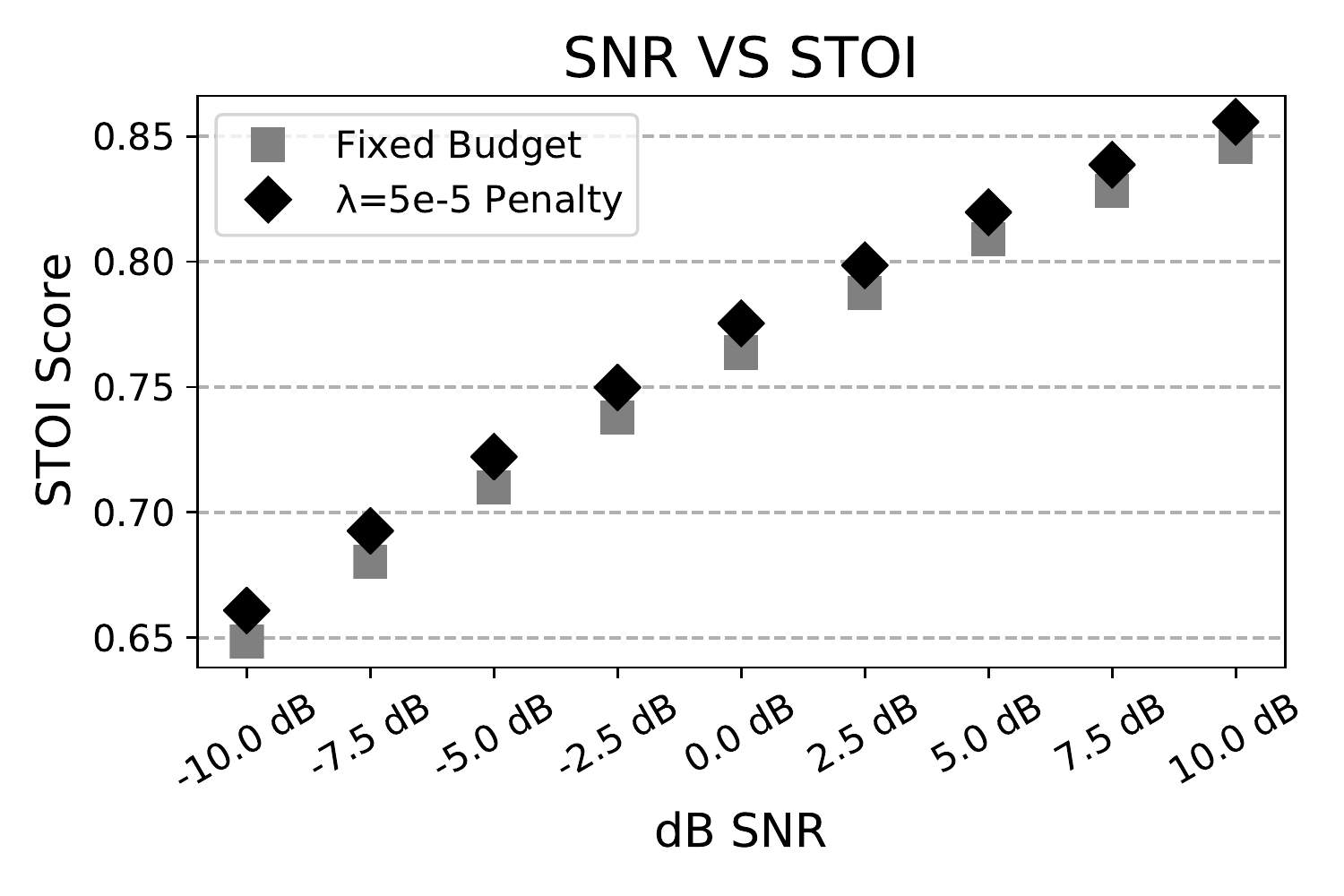}
  \caption{STOI score of the proposed model with request penalty of $5\times10^{-5}$ vs a fixed request model. Higher is better. The proposed model achieves comparable performance using less data.}
  \vspace{-1mm}
  \label{fig:snr_perf}
\endminipage
\end{figure*}

\vspace{-1mm}
\section{Experiments and Results}
\label{sec:exp_res}
\vspace{-1mm}

To evaluate the efficacy of our selection mechanism we train models with microphone request penalties ranging from $0$ to $1e-4$. The penalty of zero serves as our first baseline and verifies that requesting more data is beneficial. Then, we select the $5e-5$ penalty model and compare it to our second baseline model which requests a fixed amount of data. This experiment studies whether the selected trade-off is flexible and adapts to scene complexity.

\subsection{Training Details}
The model encoder receives two second clips of $16$kHz audio and computes STFTs with a $512$ point window, a hop of $128$, and a Hann window. The magnitude of the STFT is passed through a series of two 1D convolutional layers each with $257$ filters of size $5$ with a hop of $1$. The request network has a hidden size of $256$, a multi-head attention mechanism with $4$ heads and uses $\epsilon=10^{-4}$. The decoder has four 1D convolutional layers with the same parameters as the encoder. The final mask uses a sigmoid activation and is applied on both the real and imaginary parts of the reference channel. We train the models with the Adam optimizer~(lr $=10^{-3}$), a batch size of 16, and stop training once validation loss stops decreasing. Parameters of the encoder and decoder were tuned by hand.

\subsection{Performance Evaluation}
To evaluate the speech enhancement performance of the model, we use the short-time objective intelligibility metric~(STOI) \cite{taal2011algorithm}. We chose STOI due its simplicity. Though, other measures could be used as well. This is contrasted with the communication costs represented in seconds where costs are tracked as the average number of seconds of audio the model requested before making an enhancement prediction. Since all scenes last two seconds, a model will view at least two seconds of audio. If the model requests ten microphones, then this corresponds to it viewing $20$ seconds of audio. We track this data cost since the array is ad-hoc and getting information from a specific microphone would require transmitting it. Measuring the communication costs in seconds of audio streamed avoids any confusion about audio representation such as sampling rate, bit depth, or codec.

\subsection{Results}
\label{sec:results}

\subsubsection{Performance-Cost Trade-Off}
In this experiment, we compare the average amount of data transmitted on the test set by models which where trained with different request penalties. Intuitively, requesting to stream from all microphones throughout the duration of the scene leads to the best performance. However, some microphones are not useful and do not greatly affect performance. We verify this in Fig.~\ref{fig:tradeoff}, which shows request penalty on the x-axis and the average seconds of audio transmitted on the y-axis. As intended, when the request penalty is increased, the model requests less data. We highlight the model with no request penalty on the far left of the plot. This no-penalty model continuously streams from all microphones indicating the existence of a performance/cost trade-off.

\subsubsection{Adaptability of Selected Trade-Off}
Here, we seek to evaluate whether the performance/cost trade-off is learned in an intelligent manner. Given a communication budget, the simplest solution is to request a fixed number of microphones in all scenes. However, in simpler scenes, good performance could be achieved with fewer microphones while in more complex scenes more microphones may be necessary. In this experiment, we compare our proposed model with a request penalty of $5\times10^{-5}$ to a baseline model that always requests $18$ seconds of audio. This fixed-budget was chosen to match the performance of our proposed model in the most complex scenes. We deploy both models in scenes with a variety of SNRs, and display the amount of data requested in Fig.~\ref{fig:snr_sec} and the STOI scores in Fig.~\ref{fig:snr_perf}. The models achieve similar STOI scores despite the proposed model requesting less data.
\vspace{-1mm}
\section{Conclusion}
\label{sec:conclusion}
\vspace{-1mm}

In this work, we introduced a method for jointly-learning a microphone selection mechanism and a speech enhancement network. The attention-based microphone selection mechanism is trained to reduce communication-costs through a penalty term based on a task-performance/communication-cost trade-off. Within the trade-off, our model learns to scale with complexity, requesting less data in easier scenes and more data in harder scenes. We evaluated our model on a challenging multi-channel speech enhancement task with moving sources, reverb, and a variety of speakers and noises. The proposed model displayed performance comparable to our baseline model despite using fewer microphones. The selection mechanism described here is not specific to speech enhancement and we hope it motivates future research in neural sensor selection.

\bibliographystyle{IEEEbib}
\bibliography{strings,refs}

\end{document}